\documentclass{aa}
\usepackage{savesym}
\usepackage[varg]{txfonts}
\usepackage{euscript}
\usepackage{graphicx,bm,times,url}
\usepackage{amsmath,mathtools}
\usepackage[breaklinks, colorlinks, citecolor=blue]{hyperref}

\graphicspath{{./fig/}{./png/}}

\newcommand{\xx}{\bm{x}}
\newcommand{\ee}{\mbox{\boldmath $e$} {}}
\newcommand{\BB}{\bm{b}}
\newcommand{\AAA}{\bm{A}}
\newcommand{\m}[1]{\mathcal{#1}}
\newcommand{\healpix}{{\tt HEALPix}}
\newcommand{\planck}{{\it Planck}}

\begin{document}

\title{Is there a left-handed magnetic field in the solar neighborhood?}
\subtitle{Exploring helical magnetic fields in the interstellar medium through dust polarization power spectra}

\author{A. Bracco \inst{1} \and
        S. Candelaresi \inst{2} \and
        F. Del Sordo \inst{3} \and
        A. Brandenburg \inst{1,4,5} 
        }

\offprints{\email{andrea.bracco@su.se}}

\institute{Nordita, KTH Royal Institute of Technology and Stockholm University, Roslagstullsbacken 23, 10691 Stockholm, Sweden \and
Division of Mathematics, University of Dundee, Dundee, DD1 4HN, United Kingdom \and
Physics Department, University of Crete / FORTH, Greece \and
JILA and Laboratory for Atmospheric and Space Physics, University of Colorado, Boulder, CO 80303, USA \and
Department of Astronomy, Stockholm University, SE-10691 Stockholm, Sweden
}
             
\date{Accepted: 24 November 2018}

  \abstract
  {The analysis of the full-sky \planck\ polarization data at $850\,\mu$m revealed unexpected properties of the $E$ and $B$ modes power spectra of dust emission in the interstellar medium (ISM). The positive cross-correlations over a wide range of angular scales between the total dust intensity, $T$, with $E$ modes and, most of all, with $B$ modes has raised new questions about the physical mechanisms that affect dust polarization, such as the Galactic magnetic field structure. This is key both to better understanding ISM dynamics and to accurately describing Galactic foregrounds to the polarization of the Cosmic Microwave Background (CMB). In particular, in the quest of primordial $B$ modes of the CMB, the observed positive cross-correlation between $T$ and $B$ for interstellar dust requires further investigation towards parity-violating processes in the ISM.}
   {In this theoretical paper we investigate the possibility that the observed cross-correlations in the dust polarization power spectra, and specifically the one between $T$ and $B$, can be related to a parity-odd quantity in the ISM such as the magnetic helicity.}
{We produce synthetic dust polarization data, derived from 3D analytical toy models of density structures and helical magnetic fields, to compare with the $E$ and $B$ modes of observations. We present several models: 1) an ideal fully helical isotropic case, such as the Arnold-Beltrami-Childress field; 2) following the nowadays favored interpretation of the $T$-$E$ signal in terms of the observed alignment between the magnetic field morphology and the filamentary density structure of the diffuse ISM, we design models for helical magnetic fields wrapped around cylindrical interstellar filaments; 3) focusing on the observed $T$-$B$ correlation, we propose a new line of interpretation of the \planck\ observations advocating the presence of a large-scale helical component of the Galactic magnetic field in the solar neighborhood.}
{Our analysis shows that: I) the sign of magnetic helicity does not affect $E$ and $B$ modes for isotropic magnetic-field configurations; II) helical magnetic fields threading interstellar filaments cannot reproduce the \planck\ results; III) a weak helical left-handed magnetic field structure in the solar neighborhood may explain the $T$-$B$ correlation seen in the \planck\ data. Such magnetic-field configuration would also account for the observed large-scale $T$-$E$ correlation.}
{This work suggests a new perspective for the interpretation of the dust polarization power spectra, which  supports the imprint of a large-scale structure of the Galactic magnetic field in the solar neighborhood.}
   {}

   \authorrunning{A. Bracco et al.}
   \titlerunning{Helical magnetic fields in the ISM and dust polarization power spectra}
   \maketitle
\section{Introduction}\label{sec:intro}

Recent analyses of the sub-millimeter emission observed with the {\planck}\footnote{\planck\ (\url{http://www.esa.int/Planck}) is a project of
the European Space Agency (ESA) with instruments provided by two scientific
consortia funded by ESA member states and led by Principal Investigators from
France and Italy, telescope reflectors provided through a collaboration between
ESA and a scientific consortium led and funded by Denmark, and additional
contributions from NASA (USA).} satellite \citep{PlanckI2016} showed that the linearly polarized light of Galactic interstellar dust is an unavoidable foreground for detecting the imprint of primordial gravitational waves on the polarization of the Cosmic Microwave Background (CMB) \citep[i.e.,][hereafter PIPXXX]{BicepPlanck,planck2014-XXX}. This  discovery would represent an indirect proof of the paradigm of cosmic inflation in the early Universe \citep[e.g.][]{Hu1997}. In order to reach such tremendous achievement, an accurate model of the Galactic polarized emission is required. However, despite being discovered in the middle of the XXth century with the first starlight polarization measurements \citep{Hiltner1949,Hall1949}, because of the complexity and the variety of physical processes at play, a benchmark model for the polarization of Galactic dust is still missing. 
The acknowledged mechanism 
responsible for dust polarization can be summarized as follows: due to their asymmetric-elongated shape, spinning velocities, size-distribution, composition, and optical properties, 
large interstellar grains, from $\mu$m to mm size, 
tend to align their axis of maximal inertia (the shortest axis) with the ambient magnetic field in the interstellar medium (ISM) \citep{Chandra1953} under the action of mechanical/radiative/magnetic torques 
\citep[e.g.,][and references therein]{Davis1951,Lazarian2007,HoangLazarian2016,HoangLazarian2018}. In such a way, dust is able to emit thermal radiation with a polarization vector preferentially perpendicular to the local orientation of the interstellar magnetic field.     
Since dust grains are mixed with interstellar gas, dust
polarization observations are considered a suitable probe of the physical
coupling between the gas dynamics and the magnetic field
structure, giving insight into magnetohydrodynamical (MHD) turbulence in the ISM over a broad range of length scales, from large scales of a few 100 pc in the diffuse medium down to the sub-pc scale within molecular clouds \citep{Armstrong1995,Chepurnov2010,Brandenburg2013}. The study of dust polarization is thus an important bridge between the analysis of cosmological foregrounds and a better understanding of ISM dynamics.

The standard orthogonal base to describe a linear polarized signal is that of the Stokes parameters $I$, $Q$, and $U$. Any linear polarization can also be decomposed into two rotationally invariant quantities, which are directly derivable from the observed Stokes parameters and correspond to the {\it parity-even} $E$ modes and {\it parity-odd} $B$ modes.  
The $E$-$B$ mode decomposition is ideal to study polarization power spectra as $E$ and $B$ modes are scalar and pseudo-scalar quantities, respectively \citep{Zaldarriaga1997}. 
This decomposition was first introduced to characterize the CMB polarization, as inflationary gravitational waves would produce a well known shape to the $B$-mode power spectrum of the CMB \citep{Kamionkowski1997,Seljak1997}, and it was also applied to radio-synchrotron polarization data \citep[e.g.,][and references therein]{Robitaille2017}. However, it represents a novel technique in the case of dust polarization. 
Only thanks to the first maps of polarized emission obtained with \planck\ \citep{PIPXIX2015} it is now possible to access the full-sky statistics to explore the link between $E$-$B$ modes and ISM physics probed by dust emission at $353$~GHz ($850$~$\mu$m), whilst extrapolating dust emission properties to different frequencies might be delicate and require some corrections \citep{TassisPavlidou2015}.

As first reported in PIPXXX, and more recently confirmed and extended by \citet[][hereafter PIPXI]{PIPXI2018} using the latest version of the \planck\ data, the power-spectrum analysis of the high-Galactic-latitude sky at $353$~GHz on average showed three main results: (i) there is twice as much power in $E$ than in $B$ modes; (ii) a positive correlation between the total intensity (Stokes $I$ referred to as $T$ in the aforementioned papers) and the $E$-mode powers over a wide range of angular scales in the sky (for multipoles $l > 5$); (iii) a hint of a positive correlation between the Stokes $I$ and the $B$-mode powers, increasing at large angular scales (see Fig.~6 in PIPXI).  
These new results, not predicted by past models of dust polarization and interstellar MHD turbulence \citep[][and references therein]{Caldwell2017}, have recently been driving theoretical and numerical works to interpret the link between ISM physics and $E$-$B$ modes of dust polarization.

While \citet{Kritsuk2017} and \citet{Kandel2017,Kandel2018} claimed that the aforementioned results (i) and (ii) could be interpreted in terms of sub-Alfv\'enic MHD turbulence at high Galactic latitude (with an Alfv\'enic Mach number $M_{\rm A} < 0.5$), \citet{Caldwell2017} concluded that, because of the narrow range of theoretical parameters in their MHD simulations that could account for the observations, it is likely that \planck\ results connect to the large-scale physics that drives ISM turbulence instead of MHD turbulence itself.

If an indisputable theoretical explanation of the above results is yet to be achieved, additional observational results suggest that the $E$-to-$B$ power ratio and the $T$-$E$ correlation may be partly explained by the overall correlation between the magnetic-field morphology and the distribution of filamentary matter-density structures observed with dust emission \citep{PIPXXXII,planck2015-XXXVIII}. In particular, as suggested by \citet[][hereafter Z01]{Zaldarriaga2001}, the alignment observed at high Galactic latitudes between the structure of matter, encoded in the dust intensity, and the structure of the magnetic field, inferred from dust polarization, may be responsible for the larger $E$-mode power compared to the $B$ modes and the positive correlation between $T$ and $E$, at least on angular scales typical of interstellar filaments (for multipoles $l > 50$).
However, the alignment between filamentary density structures and magnetic fields in the ISM would struggle to answer two key questions raised by the latest \planck\ results: why does the $T$-$E$ correlation extend to very large angular scales? Where does the aforementioned result (iii), i.e., the $T$-$B$ positive correlation, come from?

In this paper we explore new ideas that may give insight into the theoretical explanation of the dust polarization power spectra. As both the temperature and the $E$-modes have opposite parity to the $B$-modes, the cross-correlations between $T$-$B$ and $E$-$B$ are expected to vanish in the absence of parity violation \citep{Zaldarriaga1997,Grain2012}. Thus, the hint of a positive $T$-$B$ correlation in the {\it Planck} data suggests the presence of a parity-breaking mechanism in the ISM, provided it is not related to residual unknown systematic errors. Since the polarization power spectra of Galactic dust emission depend on the structure of the interstellar magnetic field, we specifically investigate the impact of another pseudo-scalar quantity, namely the magnetic helicity, on the observed spectra. Magnetic helicity in the primordial Universe was also proposed to predict a non-zero correlation between CMB temperature and $B$-mode polarization fluctuations \citep{Kahniashvili2014}. This is the first attempt to explore the role of magnetic helicity on the polarization emitted by interstellar dust grains in the Milky Way. 
 
Conservation of magnetic helicity is recognized as a
key constraint on the evolution of cosmic magnetic fields, especially
those produced by large-scale dynamo action, such as the Galactic magnetic field
\citep{SSSB06,Sur2007}. The conservation of magnetic helicity also guarantees that there should be helicity fluxes across scales \citep[e.g.][]{Vishniac2001}, where small-scale magnetic turbulence would potentially drive and sustain the dynamics
of large-scale dynamo configurations \citep{BS05c}.
Thus, magnetic helicity is expected to play a role in the turbulent ISM across a broad range of scales.
 
Regarding the effect of magnetic helicity on dust polarization power spectra, our first intuition is conservative within the interpretation frame of the correlation between density filaments and the magnetic field in the ISM. The existence of helical magnetic fields wrapped around the main axis of molecular filaments has been observed \citep{Bally1987,Matthews2002,Poidevin2011,Tahani2018}, and is suggested to regulate the dynamics of such clouds against gravitational fragmentation \citep{Fiege00,Toci2015}.

We first investigate the possibility that helical magnetic fields may also thread filaments in the diffuse ISM producing the $T$-$B$ signal. Second, we propose a new perspective to interpret the dust polarization power spectra, which, in line with \citet{Caldwell2017}, suggests that the large-scale structure of the Galactic magnetic field in the solar neighborhood may partly explain both results (ii) and (iii), giving a first interpretation of the observed $T$-$B$ correlation.

The paper is organized as follows: in Sect.~\ref{sec:methods} we present the methodology employed to produce synthetic observations of dust polarization from 3D analytical models of helical magnetic field and density structures. In Sect.~\ref{sec:results} we show the results of the $E$-$B$ decomposition for three models: purely helical magnetic fields (Arnold-Beltrami-Childress, $\m{ABC}$, model); helical magnetic fields around cylindrical interstellar filaments; helical magnetic fields at large scales in the solar neighborhood. In Sect.~\ref{sec:disc} we discuss our results. The summary of the paper is presented in Sect.~\ref{sec:conc}. The paper also consists of Appendix~\ref{app:eb}, where we detail the validation of the algorithms used to compute $E$ and $B$ modes in the small-scale limit.
\section{Methods}\label{sec:methods}
We study stationary MHD models 
in a box of size $(2\pi)^3$ with periodic boundary conditions.
In order to compute and control the magnetic helicity in our theoretical 
experiments, we produce analytical models of the vector potential, $\AAA$, in Cartesian coordinates ($\ee_{x},\ee_{y},\ee_{z}$). We thus derive the following quantities:
the divergence-free magnetic field\footnote{For clarity we refer to the magnetic-field vector with the symbol "$\BB$" in order to avoid confusion with the notation of $B$-modes.}, $\BB= \nabla \times \AAA$, the total magnetic helicity, $H = \int_{V} \AAA \cdot \BB \,{\rm d}V$, and the magnetic helicity column, $\m{H} = \int_{s} \AAA \cdot \BB \,{\rm d}s$, where $s$ is any given line of sight (LOS). Since our aim is to investigate results (ii) and (iii) presented in Sect.~\ref{sec:intro}, given the magnetic field, $\BB=(b_x(x,y,z),b_y(x,y,z),b_z(x,y,z))$, and a density field, $\rho(x,y,z)$, the approach of this work is to build synthetic observations of dust polarization from which we derive $E$ and $B$ modes, as described in the next sections. 
\subsection{Synthetic observations of dust polarization}\label{sec:syntobs}

We build maps of the Stokes parameters $I$, $Q$, and $U$ adapting Eqs.~(5)--(7) of \citet{PIPXX2015}, under the assumption of optically-thin emission of dust, at least at the wavelengths observed with {\planck}. We remind the reader that from the Stokes parameters, two are the main polarization observables: the polarization fraction, $p = \sqrt{Q^2+U^2}/I$, and the polarization angle, $\psi = 0.5\arctan{(U/Q)}$, which indicates the orientation of the polarization vector projected on the plane of the sky (POS) normal to the LOS. In the case of dust polarized emission $\psi$ represents the perpendicular orientation to the magnetic-field component on the POS.

The maps are obtained integrating the cubes of the density field, $\rho$, and the magnetic field, $\BB$, along the $y$-axis, as follows  
\begin{align}\label{eq:stokes}
I &= \int_s \rho\left[1-p_0\left(\cos^2{\gamma}-\frac{2}{3}\right)\right]\, \mathrm{d} s \nonumber \\
 & \approx \sum_y \rho\left[1-p_0\left(\frac{b_x^2+b_z^2}{|\BB|^2}-\frac{2}{3}\right)\right] \Delta{y} \nonumber\\
Q &=\int_s p_0 \rho \cos{2\psi}\cos^2{\gamma}\,\mathrm{d}s \approx \sum_y p_0\rho\frac{b_x^2-b_z^2}{|\BB|^2}\Delta{y} \\
U &=\int_s p_0 \rho \sin{2\psi}\cos^2{\gamma}\,\mathrm{d}s \approx \sum_y 2p_0\rho \frac{b_x b_z}{|\BB|^2}\Delta{y}, \nonumber
\end{align}
where $p_0$ is the intrinsic polarization fraction of dust emission assumed to be constant and homogeneous across the cubes; $\gamma$ the angle between $\BB$ and the POS, and d$s$ represents the increment along the line of integration. The angles $\gamma$ and $\psi$ are defined with respect to the direction of the normal vector along the $z$-axis.    
\subsection{E-B mode decomposition}\label{ssec:EB}
While Stokes~$I$ is invariant under rotation, the Stokes~$Q$ and $U$ are not. Following \citet{Zaldarriaga1997} they transform as 
\begin{equation}\label{eq:QUstokes}
(Q +iU)'({{\bm n}}) = e^{\mp 2i\beta}(Q+iU)({{\bm n}}), 
\end{equation}
where ${{\bm n}}$ is the position in the sky and $\beta$ is the rotation of the POS reference $(\ee_1, \ee_2)$ in $\ee'_1 = \cos{\beta}\, \ee_1 + \sin{\beta}\, \ee_2$ and $\ee'_2 = -\sin{\beta}\,\ee_1 + \cos{\beta}\, \ee_2$. Notice that in the following Stokes~$I$ will be alternatively referred to as $T({{\bm n}})$ for consistency with previous works. The authors of the aforementioned paper expand these quantities in the appropriate spin-weighted basis (spherical harmonics) as 
\begin{align}\label{eq:spharm}
T({{\bm n}}) &= \sum_{lm} a_{T,lm}Y_{lm}({{\bm n}}), \nonumber\\
(Q+iU)({{\bm n}}) &= \sum_{lm} a_{2,lm} \,\prescript{}{2}Y_{lm}({{\bm n}}), \\
(Q-iU)({{\bm n}}) &= \sum_{lm} a_{-2,lm} \,\prescript{}{-2}Y_{lm}({{\bm n}}), \nonumber
\end{align}
and use the spin-raising (lowering) operators, $\eth$ ( $\overline{\eth}$ ), in order to get two rotationally-invariant quantities   
\begin{align}\label{eq:spharmrot}
\overline{\eth}^2(Q+iU)({{\bm n}}) &= \sum_{lm} \left [ \frac{(l+2)!}{(l-2)!} \right ]^{1/2} a_{2,lm}Y_{lm}({{\bm n}}), \\
\eth^2(Q-iU)({{\bm n}}) &= \sum_{lm} \left [ \frac{(l+2)!}{(l-2)!} \right ]^{1/2} a_{-2,lm}Y_{lm}({{\bm n}}). \nonumber
\end{align}
From Eq.~(\ref{eq:spharmrot}), the expansion coefficients are
\begin{align}\label{eq:coeff}
a_{T,lm} &=\int Y^{\ast}_{lm}({\bm n})T({\bm n}){\rm d}\Omega, \nonumber\\ 
a_{2,lm} &= \left [ \frac{(l+2)!}{(l-2)!} \right ]^{-1/2}\int Y^{\ast}_{lm}( {\bm n})\overline{\eth}^2(Q+iU)({{\bm n}}), \\
a_{-2,lm} &= \left [ \frac{(l+2)!}{(l-2)!} \right ]^{-1/2}\int Y^{\ast}_{lm}({\bm n}){\eth}^2(Q-iU)({{\bm n}}), \nonumber
\end{align}
which can be linearly combined into
\begin{align}\label{eq:coeffeb}
a_{E,lm} &= -(a_{2,lm}+a_{-2,lm})/2,\\
a_{B,lm} &= i(a_{2,lm}-a_{-2,lm})/2.\nonumber
\end{align}
The $E$ and $B$ modes, scalar and pseudo-scalar fields respectively, are defined as 
\begin{align}\label{eq:eb}
E({{\bm n}}) &= \sum_{lm} a_{E,lm}Y_{lm}({{\bm n}})\\
B({{\bm n}}) &= \sum_{lm} a_{B,lm}Y_{lm}({{\bm n}}).\nonumber
\end{align}
These two quantities are rotationally invariant. However, under parity transformation (i.e., changing the sign of the $x$ axis only), they behave differently. Since $Q'({{\bm n}'})=Q({{\bm n}})$ and $U'({{\bm n}'})=-U({{\bm n}})$, from Eqs.~(\ref{eq:coeff}) and (\ref{eq:coeffeb}), one can show that $E'({{\bm n}'})=E({{\bm n}})$ while $B'({{\bm n}'})=-B({{\bm n}})$. Thereby, $E$ and $B$ modes are even and odd quantities, respectively, under parity transformations. Such property makes them interesting quantities to explore the link between dust polarization and magnetic helicity, which changes sign as well under parity transformation.

The usual statistical description of the three scalar/pseudo-scalar quantities defined above ($T,E,\,{\rm and}\, B$) is based on their power spectra 
\begin{equation}\label{eq:cl}
C^{XY}_{l} = \frac{1}{2l+1}\sum_{m}\langle a^{\ast}_{X,lm} a_{Y,lm} \rangle,
\end{equation}
auto ($X=Y$) and crossed ($X \neq Y$), where $X$ and $Y$ may refer to $T$, $E$, or $B$. We also use the normalized parameter introduced by \citet{Caldwell2017} to quantify the correlation among the power spectra  
\begin{eqnarray}\label{eq:rpar}
r^{XY}_l = \frac{C^{XY}_{l}}{\sqrt{C^{XX}_l \times C^{YY}_l}},
\end{eqnarray}
so that in case of perfect positive (negative) correlation $r^{XY}_{l} = 1\,(-1)$, and in case of absence of correlation $r^{XY}_{l} = 0$.

In this paper we compute both $E$-$B$ modes on the sphere, as described in the present section, and 
their small-scale
limit in 2D maps (see Appendix~\ref{app:eb} for details). 
In the case of 2D maps there is no loss of generality, although some concerns about the boundary conditions are required. Moreover the multipole $l$ is replaced by the wavenumber $k$, and the coefficients of the spherical harmonics are replaced with the Fourier transform coefficients. 
\begin{figure*}[htbp] 
   \centering
   \includegraphics[width=\textwidth]{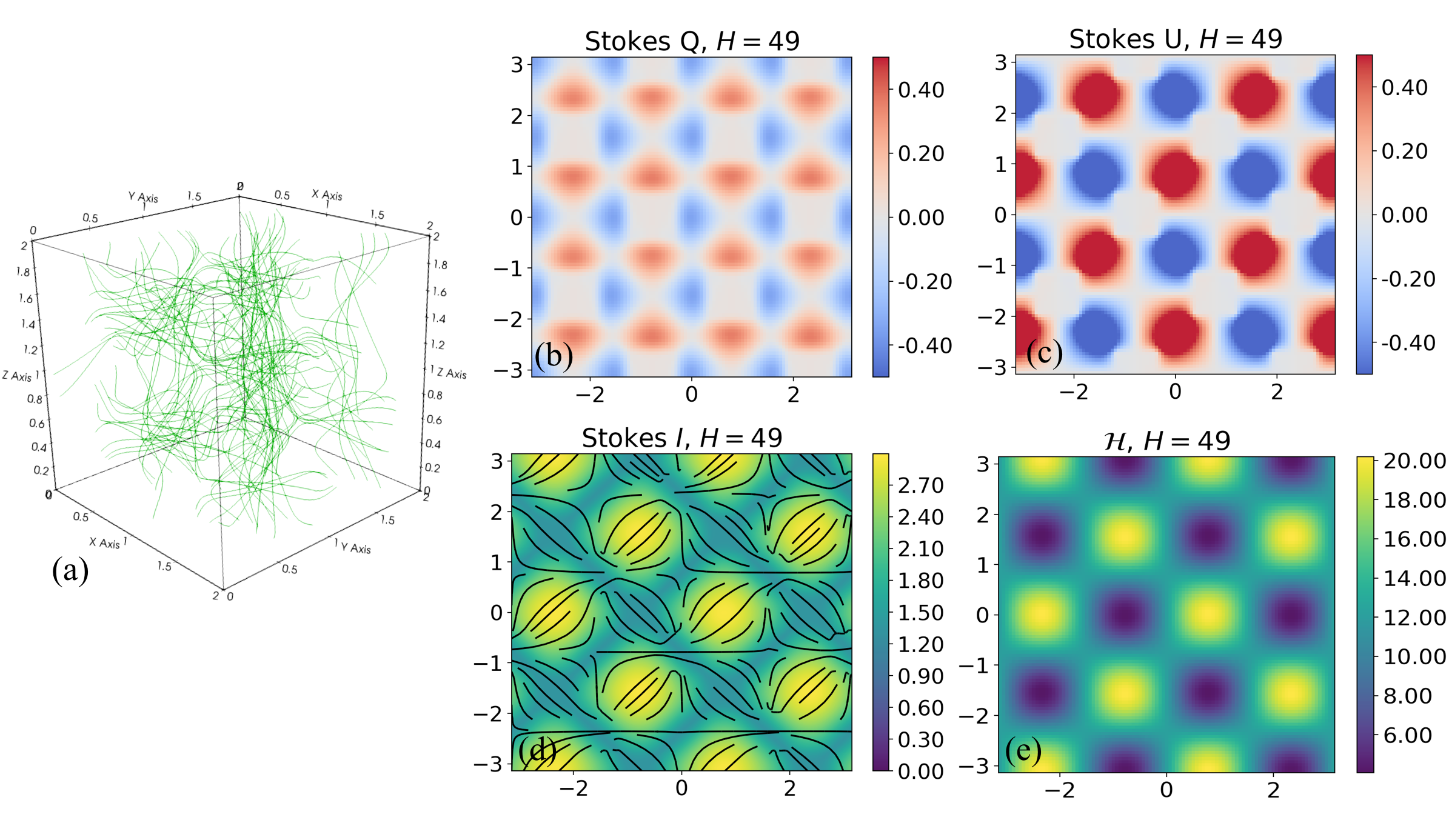}
   \caption{Helical magnetic field, ($\m{ABC}$) model, with $\m{A}$=$\m{B}$=$\m{C}$=1 and $\lambda=+2$. The total helicity, integrated over the cube, is $H=49$, in normalized units. The box has a uniform density field. From left to right: 3D rendering of the magnetic-field lines in green ({\it panel a}); projected map of Stokes $Q$ with line-of-sight integration along the y-axis ({\it panel b}); projected map of Stokes $U$ ({\it panel c}); projected map of Stokes $I$ with overlaid magnetic-field lines tracing the orientation of $\BB$ on the plane of the sky ({\it panel d}); helicity-column map $\m{H}$ ({\it panel e}).}
   \label{fig:abc1}
\end{figure*}
In order to perform the power-spectra analysis with spherical harmonics on the sphere we make use of the {\it{healpy.sphtfunc.anafast.py}} routine of the \healpix\footnote{http://healpix.jpl.nasa.gov} healpy package.
All the codes written in Python used for this analysis can be accessed and downloaded from the {\it HBEB} GitHub page: \url{http://github.com/abracco/cosmicodes/tree/master/HBEB}.   
\section{Models and results}\label{sec:results}
In this section we present three models with distinct magnetic field and density configurations. We discuss the respective results applying the methodology described in Sect.~\ref{sec:methods}. 
\begin{figure*}[htbp] 
   \centering
   \includegraphics[width=\textwidth]{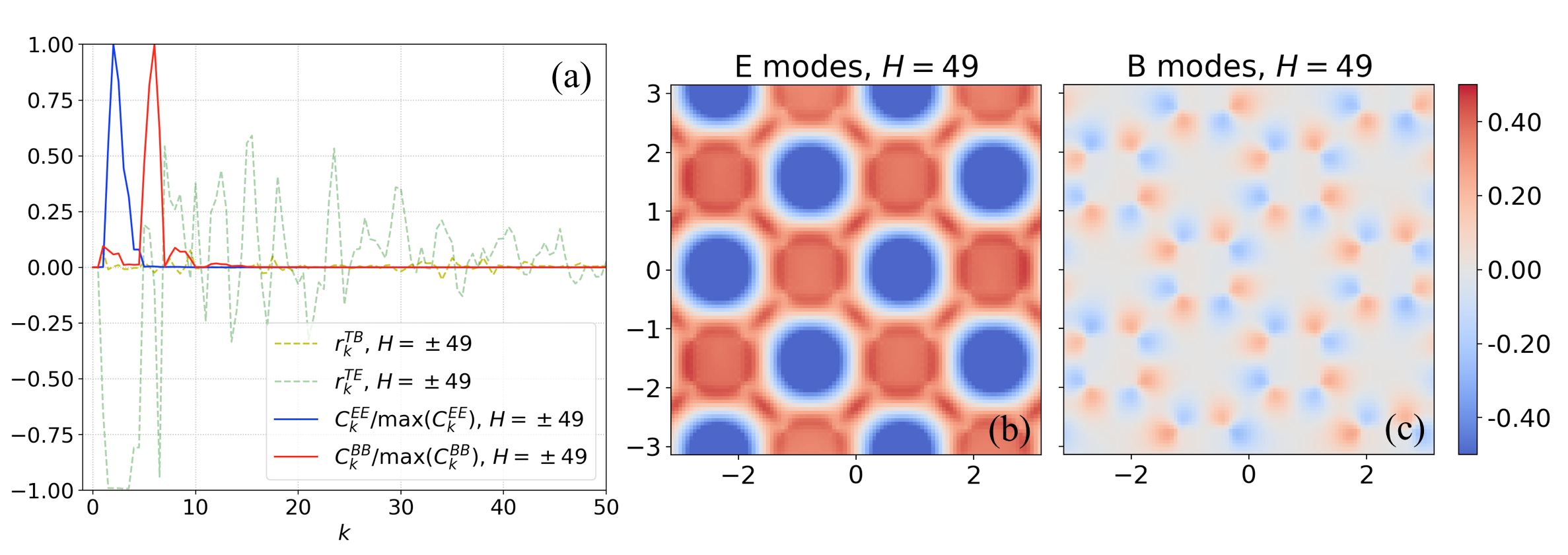}
   \caption{$E$ and $B$ modes for the case of $\m{ABC}$ field illustrated in Fig.~\ref{fig:abc1}. In {\it panel a} we show the normalized autocorrelation power spectra of $E$ and $B$ modes, where max$(C^{EE}_{k})$/max$(C^{BB}_{k}) \approx 60$, and the parameters $r^{TE}_{k}$ and $r^{TB}_{k}$ (see Eq.~(\ref{eq:rpar})). These functions attain the exact same value independently of the sign of the total helicity, $H$.  {\it Panels b} and {\it c} show the maps of $E$ modes and $B$ modes with line-of-sight integration along the $y$-axis.}
   \label{fig:abc2}
\end{figure*}
\begin{figure*}[htbp] 
   \centering
   \includegraphics[width=\textwidth]{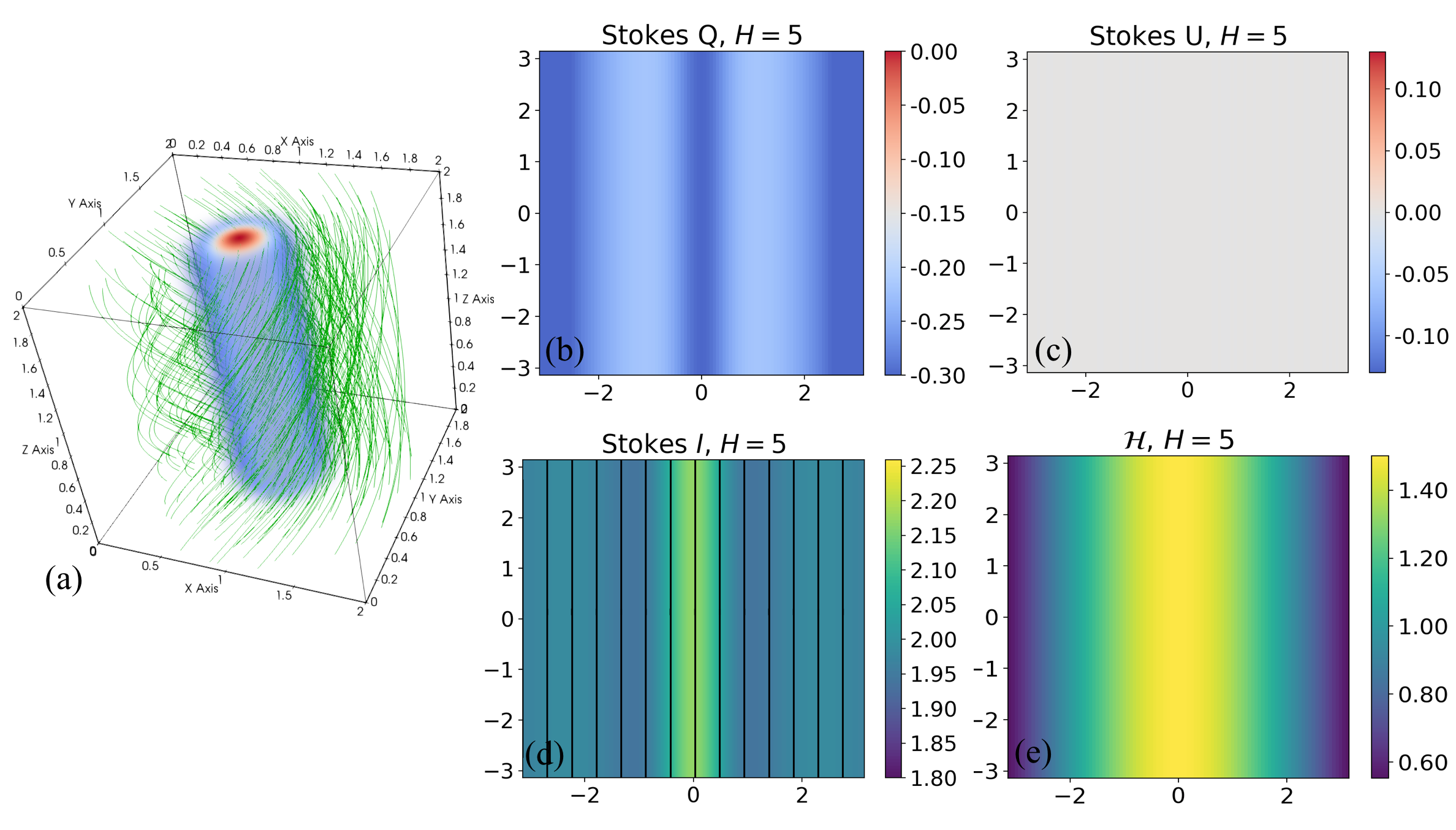}
   \caption{Same as in Fig.~\ref{fig:abc1} for the helical magnetic field model wrapped around a cylindrical density structure that mimics a filament. The model has $\alpha_b = +1$ and total helicity $H=+5$. Despite the 3D helical structure of the magnetic field, the projected pattern does not show any inclined field component with respect to the main axis of the filamentary structure (see {\it panels c} and {\it d}).}
   \label{fig:fil1}
\end{figure*}
\subsection{Arnold-Beltrami-Childress ($\m{ABC}$) model}\label{ssec:ABC}
The first case that we take into account is a fully helical magnetic field in a box with homogeneous density ($\rho=1$). The vector potential of our field is that of an $\m{ABC}$ flow \citep{Galloway1986}: 
\begin{align}\label{eq:abc}
A_{x} &= {\m{A}}\sin{(\lambda z)} + {\m{C}}\cos{(\lambda y)},\nonumber \\
A_{y} &= \m{B}\sin{(\lambda x)} + \m{A}\cos{(\lambda z)}, \\ 
A_{z} &= \m{C}\sin{(\lambda y)} + \m{B}\cos{(\lambda x)}, \nonumber
\end{align}
where $\m{A}$, $\m{B}$, $\m{C}$, and $\lambda$ are scalars.
The $\m{ABC}$ flow is fully helical, so that the magnetic-field vector and the total magnetic helicity are given by, 
\begin{align}\label{eq:BH_abc}
\BB &= \lambda \AAA, \\
H &= \lambda\, \int_{V} |{\AAA(\lambda)}|^2 \,{\rm d}V. \nonumber
\end{align}
This field is highly symmetric and isotropic when $A=B=C$. First, we focus on the case with $A=B=C=1$ and $\lambda = +2$, which produces a right-handed helical magnetic field with $H = 49$ (in normalized units, see Fig.~\ref{fig:abc1}). The Stokes $Q$, $U$, and $I$ show very regular patterns in the projected maps. Similarly, the magnetic helicity column, $\m{H}$, shows a periodic behavior that overlaps with Stokes $I$, in which the structure seen in the map only depends on the magnetic-field geometry encoded in Eq.~(\ref{eq:stokes}). In this case $p_0=1$.
Although this model is a purely theoretical and unphysical experiment, it allows us to start exploring the impact of magnetic helicity on dust polarization power spectra. The first result is that in this configuration, despite the specific value of the parameters at play, we find that the correlation between $T$ and $B$-modes, as well as with $E$-modes, does not change when $\lambda$, thus $H$, changes sign. The same holds for the autocorrelation between $E$ and $B$ modes. As shown in Fig.~\ref{fig:abc2}, $r_k^{TB}$, $r_k^{TE}$, $C_k^{EE}$, $C_k^{BB}$ attain the same values that only depend on the modulus of $H$ but not on its sign. This result is the same regardless of the choice of $\m{A}$, $\m{B}$, $\m{C}$, and $\lambda$.  

\subsection{Helical magnetic fields around interstellar filaments}\label{ssec:hlx}

The $\m{ABC}$ test suggests that, in case of isotropic magnetic-field configurations, the sign of $H$ does not play any role in the dust polarization power spectra. However, in the case of interstellar filamentary structures, which is the relevant case we want to explore, a clear source of anisotropy is provided by the main axis of the filaments itself. Thus, in this section we develop a more realistic toy model that describes a matter-density structure in the form of a cylindrical filament with a helical magnetic-field wrapped around it. We model the filament with an axisymmetric density profile given by
\begin{eqnarray}\label{eq:density}
\rho(\xx) = (\rho_i -\rho_e)e^{-\frac{x^2+y^2}{r^2_0}}+\rho_e,
\end{eqnarray}
where $r_0$ is the radial extent of the filament, $\rho_i$ is the matter density at its center, and $\rho_e$ is the external density.
\begin{figure*}[htbp] 
   \centering
   \includegraphics[width=\textwidth]{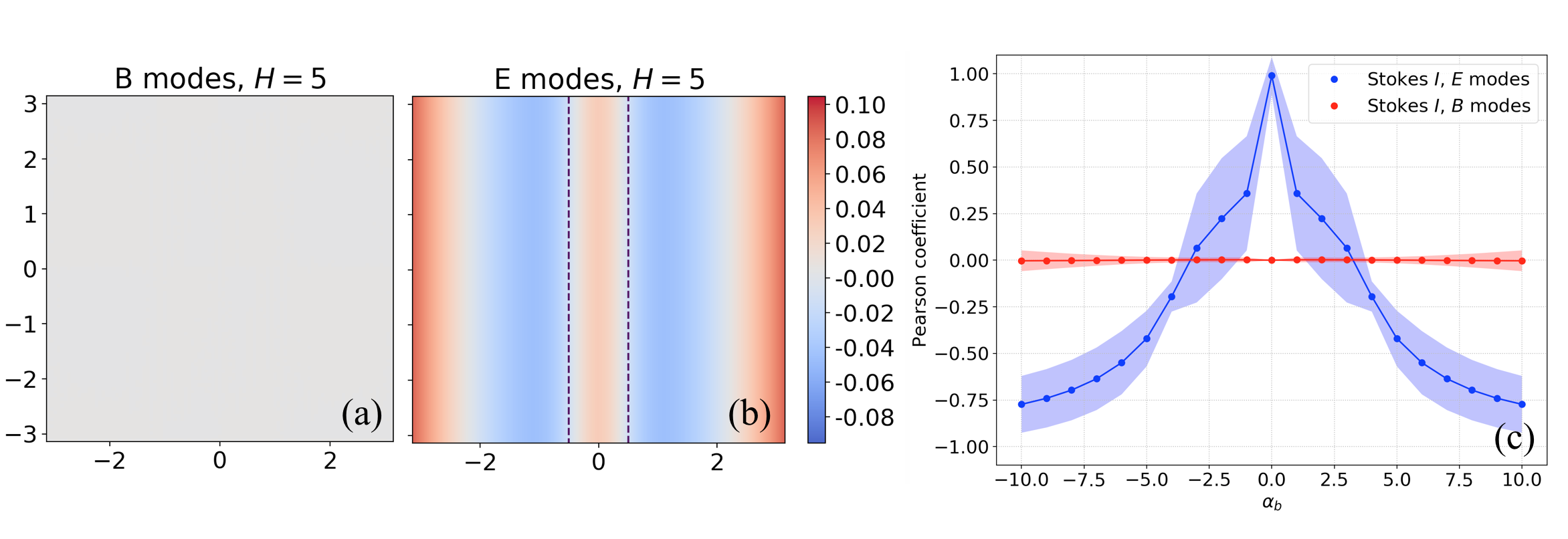}
   \caption{Maps of $B$ and $E$ modes ({\it panel a} and {\it b}, respectively) for the model shown in Fig.~\ref{fig:fil1}. {\it Panel c} shows the Pearson coefficients between Stokes~$I$ ($T$) and $E$-$B$ mode maps within the contours shown in {\it panel b}, or the 5\% brightest pixels of Stokes~$I$. The data points show the mean of 100 random rotations of the filament-magnetic-field system with respect to the line of sight. The shadows represent the 1-$\sigma$ error of the 100 rotations. Regardless of the observer's point of view with respect to the filament, helical magnetic fields wrapped around filamentary structures do not produce any $B$-mode signal correlated with the filament in Stokes~$I$ (see main text for details).}
   \label{fig:fil2}
\end{figure*}
\begin{figure}[htbp] 
   \centering
   \includegraphics[width=0.5\textwidth]{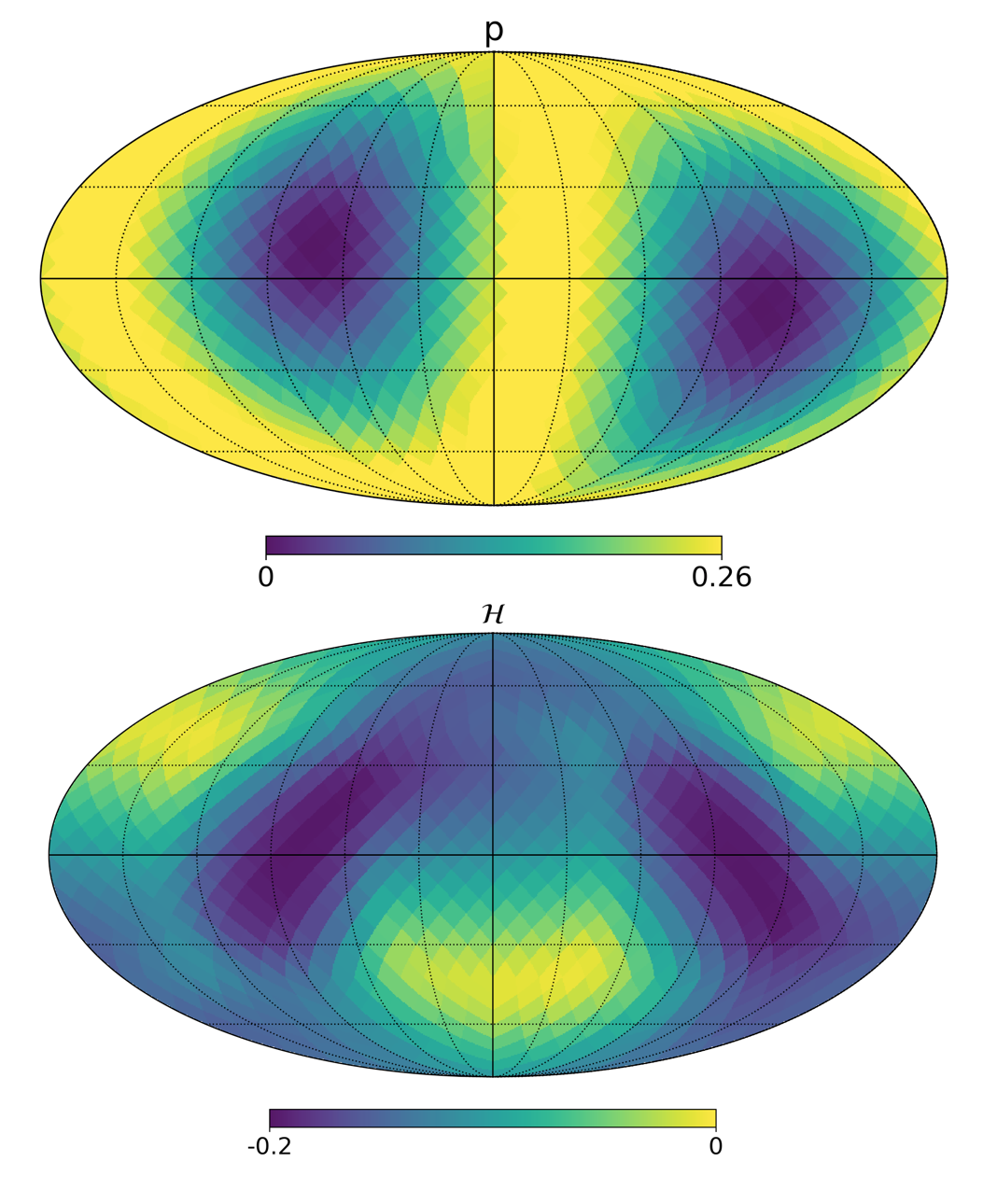}
   \caption{Mollweide projections of the polarization fraction, $p$, and magnetic-helicity column, $\m{H}$, produced placing the observer at the center of the filamentary structure described in Sect.~\ref{ssec:hlx} so to emulate the structure of the Galactic arm in the solar neighborhood. For this specific model, both the direction of the local arm and that of the poloidal component of the magnetic field is [$l^{\rm G}_0,b^{\rm G}_0$] = [$70^{\circ}$,$+10^{\circ}$], $\alpha_{b} = -20\%$, and $p_0 = 26 \%$. A Galactic coordinate grid centered in [l,b] = [$0^{\circ}$,$0^{\circ}$] and with steps of $30^{\circ}$ is superposed.}
   \label{fig:mean2}
\end{figure}

To obtain a helical magnetic field around it, we combine a uniform magnetic field, parallel to the main filament axis, and a toroidal magnetic field. The uniform magnetic field is $\BB_{\rm 0} = b_0 \ee_z$, where $b_0$ is its strength. Its vector potential is
\begin{equation}
\AAA_{\rm 0} = -\frac{yb_0}{2}\ee_x + \frac{xb_0}{2}\ee_y.
\end{equation}
The toroidal field, $\BB_{\rm t}$, is of the form
\begin{eqnarray}
\BB_{\rm t}(\xx) = 2\kappa e^{-(x^2+y^2)} (-y\ee_x + x\ee_y),
\end{eqnarray}
where $\kappa$ sets the strength of the toroidal component and the sign of $H$. We define the toroidal-to-uniform field strength as $\alpha_b=\kappa/b_0$. The vector potential of $\BB_{\rm t}$ is 
\begin{eqnarray}
\AAA_{\rm t}(\xx) = \kappa e^{-(x^2+y^2)} \ee_z.
\end{eqnarray}

The $\alpha_b$ parameter determines the large-scale configuration of the magnetic field and the pitch angle of the helix. If $\alpha_b=0$ a uniform field along the filament is generated; $\alpha_b \neq 0$ produces a right-handed (left-handed) helix wrapped around the filament if $\alpha_b > 0$ ($\alpha_b < 0$). The pitch angle depends on the distance from the center of the filament but if $\alpha_b$ assumes large absolute values the magnetic field tends to acquire a configuration perpendicular to the main filament axis.

As an example, in Fig.~\ref{fig:fil1} we show a model where the filament is oriented along the $z$-axis, $p_0 = 26\%$ \citep[in the diffuse ISM,][]{PIPXLIV2016}, and the toroidal component of the magnetic field is in equipartition with the uniform component, or $\alpha_b = +1$. The 3D rendering in {\it panel a} shows that, at large scale, the magnetic field, $\BB = \BB_0+\BB_{\rm t}$, tends to be parallel to the filament \citep{PIPXXXII,planck2015-XXXVIII} although a helical component wrapped around it appears too. Nevertheless, despite the helical magnetic-field structure, the Stokes parameter maps are those of a uniform field along the filament, where $Q < 0$ and $U = 0$ (see also the POS magnetic-field lines in {\it panel d}). Correspondingly, as expected from Z01, only $E$ modes are produced as shown by {\it panels a} and {\it b} in Fig.~\ref{fig:fil2}.

We investigate the possibility that $B$ modes appear if an angle between the filament and the LOS is introduced. In that case, the boundary conditions in the box are not periodic anymore. Instead of computing $E$ and $B$ modes, as explained in Appendix~\ref{app:eb}, we use the argument raised by Z01 for which, in the case of filaments, $E$ and $B$ modes correspond to the Stokes parameters in the reference frame of the filament itself. We rotate the Stokes $Q$ and $U$ by an angle $\beta$ (see Eq.~(\ref{eq:QUstokes})), or the projected orientation of the filament in Stokes $I$ with respect to the $z$-axis, and we obtain the rotated Stokes parameters, $Q_{\rm rot}\propto-E$ and $U_{\rm rot}\propto-B$ (see Eq.~(18) in Z01). This allows us to estimate the correlation between the filament in Stokes $I$ and its counterpart in $Q_{\rm rot}$ and/or $U_{\rm rot}$, thus in $E$ and/or $B$, respectively.

{\it Panel c} of Fig.~\ref{fig:fil2} shows the Pearson coefficients between $I$ and $-Q_{\rm rot}$ ($E$) and $-U_{\rm rot}$ ($B$) computed for the $5\%$ brightest pixels in Stokes $I$ (see dashed contours in {\it panel b} as an example). We show the Pearson coefficients as a function of $\alpha_b$. Each data point in the plot corresponds to the mean of $100$ random rotations in the box of the filament with respect to the LOS. The shades represent the $1$-$\sigma$ error of the 100 rotations. This correlation plot reveals that in our models of helical magnetic fields wrapped around filaments we are not able to reproduce any correlation between $I$ and $B$ modes, i.e., any $T$-$B$ correlation, regardless of the value of $\alpha_b$ and the viewing angle of the filament. On the other hand, Fig.~\ref{fig:fil2} shows that the correlation between $T$ and $E$ changes from positive to negative going from parallel magnetic fields along the filaments ($\alpha_b = 0$) to almost perpendicular configurations ($|\alpha_b| \gg 0 $).
\begin{figure}[htbp] 
   \centering
   \includegraphics[width=0.5\textwidth]{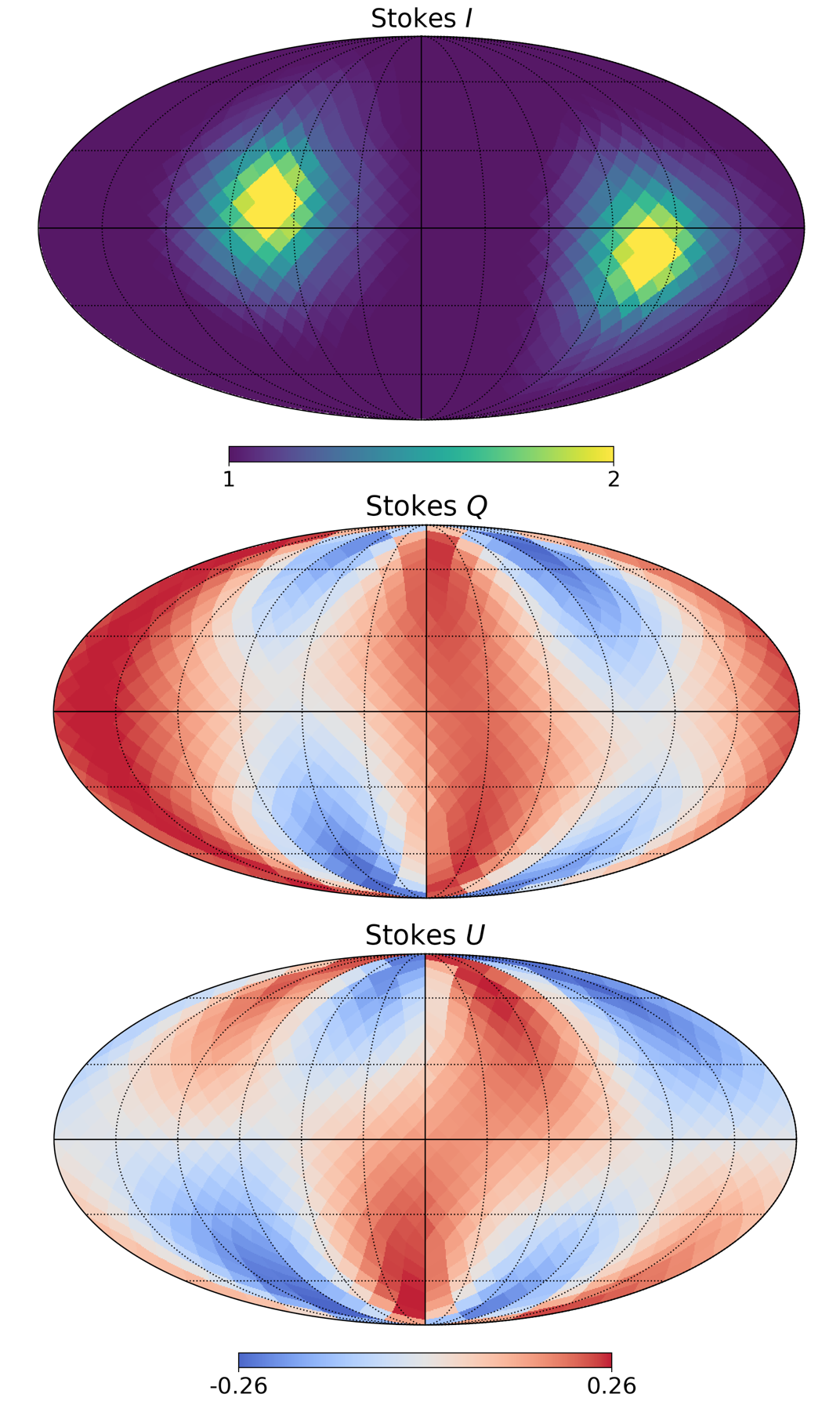}
   \caption{Mollweide projections of the Stokes parameters $I$, $Q$, and $U$ using the same model as in Fig.~\ref{fig:mean2}.}
   \label{fig:mean1}
\end{figure}
\begin{figure*}[htbp] 
   \centering
   \includegraphics[width=\textwidth]{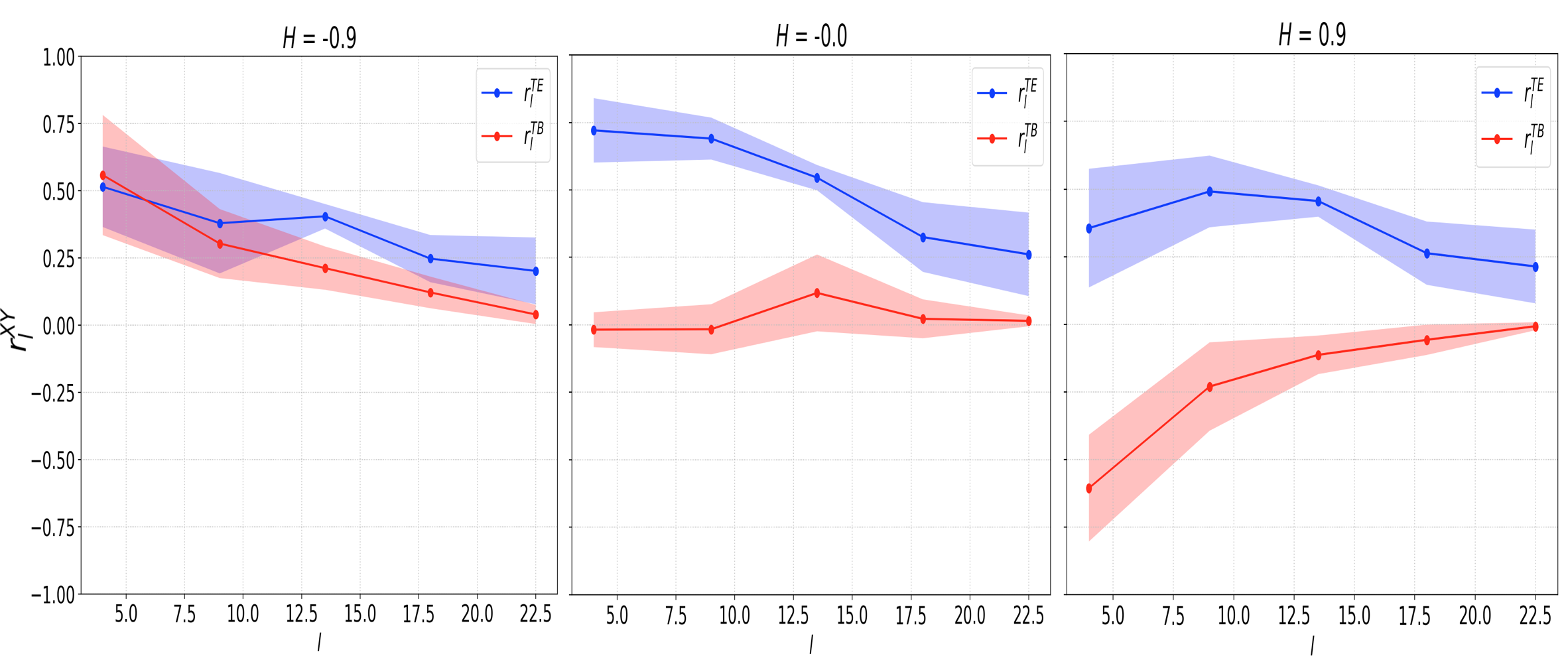}
   \caption{Mean values of the parameters $r_l^{XY}$, with $X=T$ and $Y=E,B$ (blue and red, respectively) in bins of the multipole $l$ (with 1-$\sigma$ errors in colored shades) for the model in which the observer is located within the helical structure of the magnetic field. The diagrams change from left to right for models with $\alpha_b = -20\%$, $0\%$, and $+20\%$. A weak left-handed helical magnetic field would produce a positive $T$-$B$ and $T$-$E$ correlations.}
   \label{fig:mean3}
\end{figure*}

\subsection{A new perspective: helical magnetic fields at large scale in the solar neighborhood}\label{ssec:solar}

Our simple approach, so far, seems to disfavor the interpretation of the observed $T$-$B$ correlation in the \planck\ data in terms of filamentary structures of matter in the ISM, which would be morphologically associated with the magnetic-field topology, as it has been speculated in the case of the observed $T$-$E$ correlation and the $E$-to-$B$ power ratio (see Sect.~\ref{sec:intro}). In this section we propose an alternative perspective. We do not pretend to accurately fit the data, but we rather offer a different point of view via a numerical experiment, which allows us to partly account for the observed dust polarization power spectra.

Our approach is to consider the observer within the helical structure of the magnetic field, instead of looking at it from outside. This case may be thought of as being at the position of the Sun in the Milky Way and observing with a satellite the surrounding dust polarization coming from the local spiral arm, in which we are embedded. In this configuration, the helical magnetic field would correspond to a large-scale feature of the Galactic magnetic field in the solar neighborhood, which may be controlled with the $\alpha_b$ parameter.

The case with $\alpha_b = 0$ corresponds to having only a uniform magnetic field orientation in the vicinity of the Sun. Such a uniform orientation, within a few hundred pc from us, has already been suggested by several works concerning dust polarization analyses both in extinction and in emission \citep[e.g.,][]{Heiles1996a, PIPXLIV2016, Alves2018}. We make a step forward proposing that superimposed onto the uniform magnetic field a toroidal component may exist, which would generate a non-null magnetic helicity.
A similar scenario was put forward by \cite{Mathewson68} on the basis of polarization measurements of stars within 500 pc from the Sun. 

In practice, we produce similar 3D boxes as in Sect.~\ref{ssec:hlx}, where the density cylinders now represent the local spiral arm in the Milky Way instead of interstellar filaments. At this stage, we place the observer at the center of the cube and we use Eqs.~4, 5, and 6 in \citet{PIPXLIV2016} to obtain, for each voxel of the box, the values of $\gamma$ and $\psi$. We thus transform Eqs.~\ref{eq:stokes} in spherical coordinates and, in order to get the projected Stokes parameters on the celestial sphere in Galactic coordinates, we integrate the cubes over the radial direction. This projection and the tessellation on the sphere is made using the \healpix\ healpy package with a resolution of NSIDE = 8.

In Figs.~\ref{fig:mean2} and \ref{fig:mean1} we show the full-sky maps of the polarization fraction, $p$, of $\m{H}$, and of the Stokes parameters for a model with a tentative direction of the local arm/uniform magnetic field toward Galactic coordinates $[l^{\rm G}_0,b^{\rm G}_0]=[70^{\circ},+10^{\circ}]$, $p_0=26\%$ \citep[][]{PIPXLIV2016,Alves2018}, and $\alpha_b = -20\%$. The maps clearly show the presence of the uniform magnetic field, which can best be seen in $p$, where the large regions of low polarization fraction are due to the projection factor $\cos^2{\gamma}$ in Eq.~(\ref{eq:stokes}) that becomes negligible when $\BB_0$ points along the LOS. The $\m{H}$ map, only derivable from models and not from observations, also shows curious features that correspond to $p$. The full-sky maps of $I$ ($T$), $Q$, and $U$ allow us to compute $E$ and $B$ modes, and the corresponding power spectra, on the sphere using spherical harmonics.

\begin{figure}[!htbp] 
   \centering
   \includegraphics[width=0.5\textwidth]{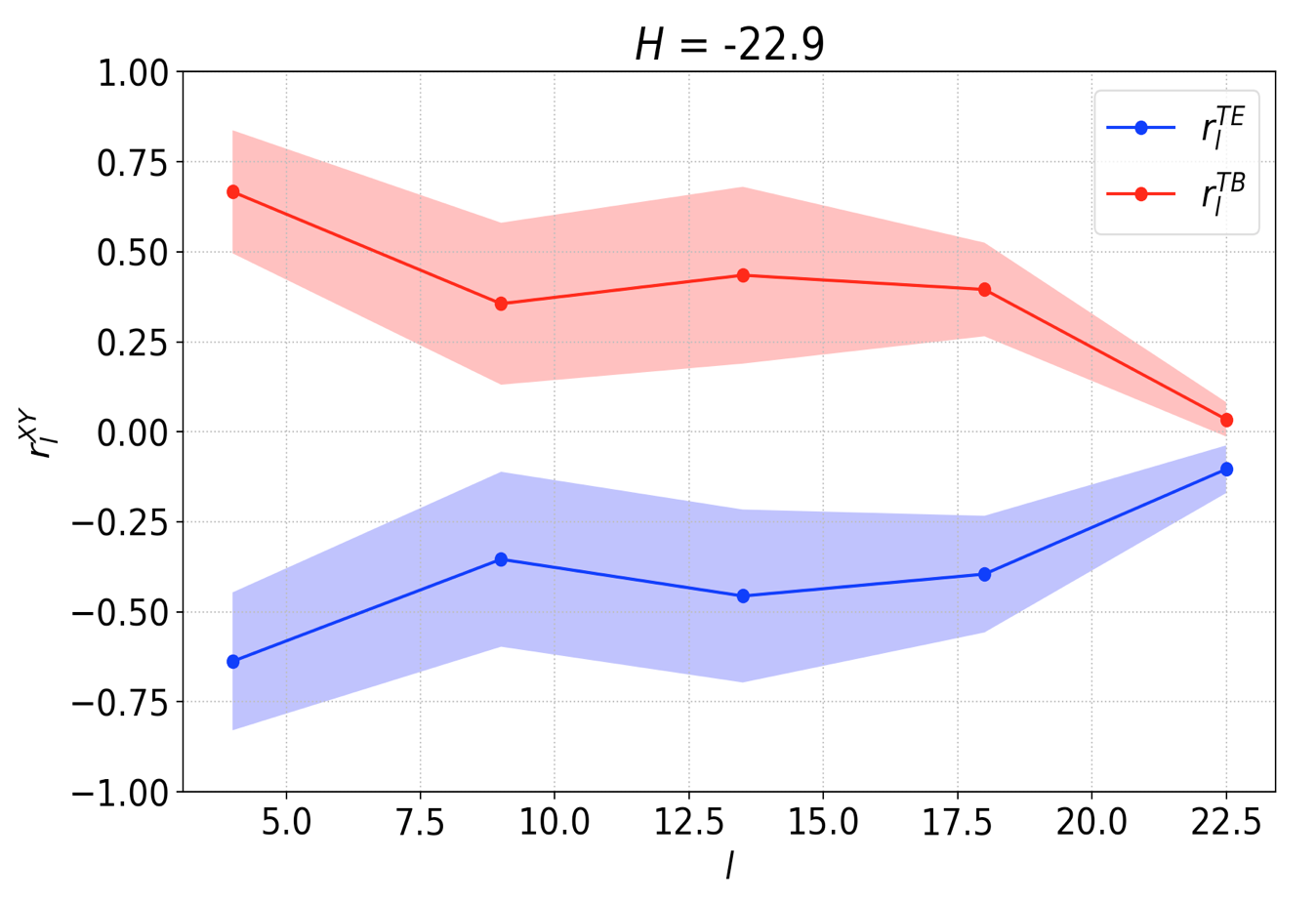}
   \caption{Same as in Fig.~\ref{fig:mean3} for $\alpha_b=-500\%$. A strong toroidal component of the magnetic field compared to the uniform component would produce a negative $T$-$E$ correlation.}
   \label{fig:mean4}
\end{figure}

In Fig.~\ref{fig:mean3} we show the mean values of $r_l^{TE}$ and $r_l^{TB}$ binned in multipoles with the corresponding 1-$\sigma$ error as a function of $\alpha_b$. The important and interesting result is that the presence of a large-scale helical component of the Galactic magnetic field may indeed generate a non-null $T$-$B$ correlation, which could not be produced by a uniform magnetic field alone (see central panel and \citet{PIPXLIV2016}). In particular, left-handed magnetic fields would reproduce the trend observed in the \planck\ data, showing a positive $T$-$B$ correlation at large angular scale, which would be negative for right-handed magnetic fields. The slight additional changes in the shape of the parameter $r^{TB}_l$ with opposite values of $H$ are caused by a large-scale asymmetry related to the mean direction of $\BB_0$, which points towards the Galactic latitude $b^{\rm G}_0 =+10^{\circ}$. The $T$-$B$ correlation depends on the absolute value of $\alpha_b$. In Fig.~\ref{fig:mean4}, we show a case where $\alpha_b = -500\%$. Having a left-handed magnetic field with a very strong toroidal component, compared to the uniform one, one would be able to produce a positive $T$-$B$ correlation, however, this would generate a negative $T$-$E$ correlation at large scale, which is not observed in the data.

\begin{figure}[!htbp] 
   \centering
   \includegraphics[width=0.5\textwidth]{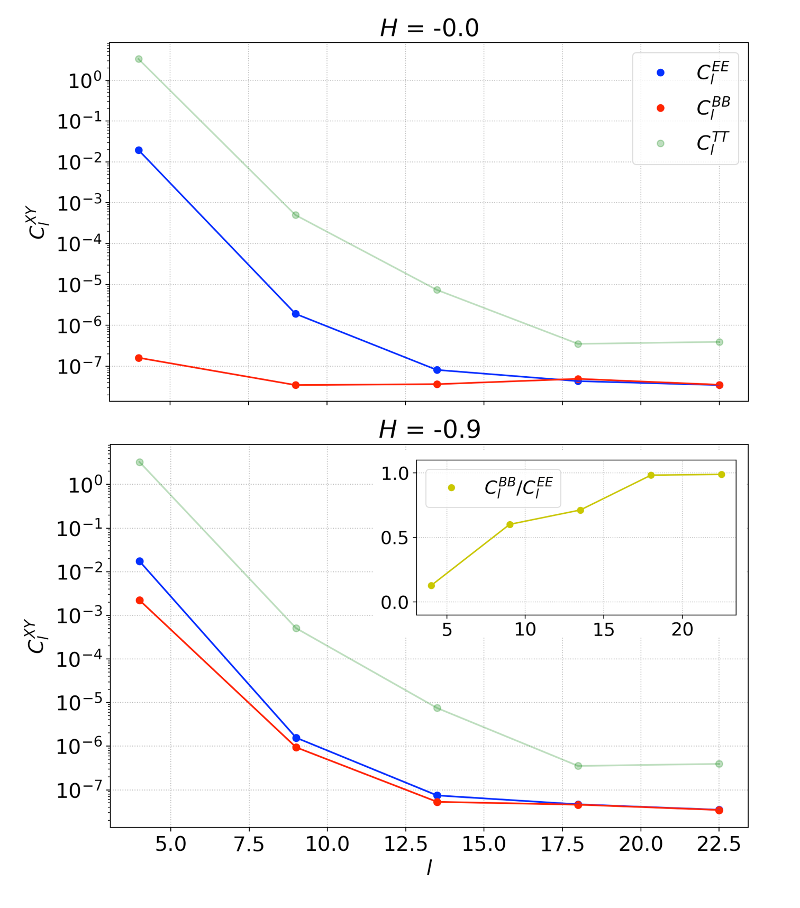}
   \caption{Polarization power spectra for $\alpha_b=0$ ({\it top}) and $\alpha_b=-20\%$ ({\it bottom}). As shown by the latter model, the relative power between $B$ and $E$ modes is not constant; see the ratio in yellow in the sub-plot in the bottom panel.}
   \label{fig:mean5}
\end{figure}

In Fig.~\ref{fig:mean5} we display the binned polarization power spectra for
$\alpha_b = 0$ and $\alpha_b = -20\%$, which show that the relative power between $E$ and $B$ modes is scale dependent and not fixed in our models.
The $C^{BB}_l/C^{EE}_l$ ratio is not shown for $\alpha_b = 0$ as $B$ modes
are not present. Although beyond the main scopes of this work, the $\alpha_b$ parameter may be tuned to reproduce the observed $E$-to-$B$ ratio at large scale. 
\section{Discussion}\label{sec:disc}
In this work we investigated the $T$-$E$ and $T$-$B$ correlations of the power spectra found in the \planck\ data. We used 3D analytical toy models of magnetic fields, where we could change the value of magnetic helicity. We produced synthetic observations from the models to qualitatively compare with the dust polarization observational results of {\planck}. A different approach based on fully-helical MHD numerical simulations is employed in a separate paper to probe the role of anisotropic turbulence in the dust-polarization power spectra \citep{Brandenburg2018}.

Using the $\m{ABC}$ flow as a reference model in Sect.~\ref{ssec:ABC}, we showed that isotropic and fully helical magnetic-field configurations cannot be probed through dust-polarization power spectra. In this case both $E$ and $B$ mode power spectra, and their cross-correlations, are completely independent on the sign of magnetic helicity.

In Sect.~\ref{ssec:hlx} we introduced a source of anisotropy in the magnetic field following the line of interpretation acknowledged for the $E$-to-$B$ power ratio and the $T$-$E$ signal (at least for multipoles $l > 50$), or the correlation in the diffuse ISM between the magnetic-field morphology and the distribution of interstellar matter organized in filaments. We produced toy models of cylindrical interstellar filaments wrapped in helical magnetic fields. We designed the magnetic field to be composed of a poloidal-uniform component along the main axis of the filaments and a toroidal component around it. On the one hand we were able to reproduce the positive $T$-$E$ correlation for weak toroidal magnetic-field components. On the other hand, none of our filamentary models, regardless of the parameters at play, enabled us to generate any $B$-mode counterpart of the density filaments. Most likely, the reason why we did not find traces of $B$ modes in these models is that, under the assumption of optically thin dust emission, any contribution that could provoke $B$ modes (i.e., a magnetic field oriented at $\pm 45^{\circ}$ with respect to the density filament) averaged out in the process of integrating the modeled cubes along the LOS. In conclusion, unless we consider an optical depth dependence of dust polarization (which is not realistic at \planck\ wavelengths), our toy models of helical magnetic fields around interstellar filaments are not able to probe the $T$-$B$ correlation observed in the \planck\ data.
On the one hand, our models may be too simplistic to capture the physics of
interstellar filaments, while on the other hand we also notice that the filamentary
density structure observed in the \planck\ maps projected on the high-latitude sky may
well be either the result of projection effects or, as probed by spectroscopic data
of atomic hydrogen \citep{Clark2015}, that of velocity crowding \citep{Lazarian2018}.

In Sect.~\ref{ssec:solar}, we showed that magnetic helicity may indeed play a role on the $T$-$B$ correlation if only the favored interpretation of the correlation between density filaments and magnetic fields was opened to a new perspective. As already speculated by \citet{Caldwell2017}, part of the signal observed by \planck\ may not come from MHD turbulent processes in the ISM but rather from a large-scale structure of the Galactic magnetic field in the solar neighborhood. We showed that a helical left-handed large-scale structure of the Galactic magnetic field around the Sun, with a weak toroidal component compared to the uniform component, may explain both the large-scale positive $T$-$E$ and, most of all, $T$-$B$ correlations found in the \planck\ data. Interestingly left-handed helical magnetic fields in the intergalactic medium were already suggested by \citet{Tashiro2014} using gamma-ray data from the {\it Fermi} satellite. The possibility of having such magnetic-field structure in the Milky Way was discussed long time ago by \citet{Fujimoto1969} in the attempt to explain contemporary starlight polarization and Faraday rotation measurements of polarized extragalactic radio sources. In the presence of magnetic flux freezing in the interstellar gas, the authors claimed that a helical magnetic field structure around the local spiral arm would have left a clear kinematic signal that one could observe. If at that time observational data of the kinematics of the diffuse ISM were rare and limited, nowadays we have access to very high-quality HI data of the full sky \citep{HI4PI2016}. Although beyond the scopes of the present work, it would be interesting for the future to investigate the HI kinematics in the quest of such a helical magnetic-field structure at large scale in the solar neighborhood.

We would like to remind the reader that although our results only allow us to qualitatively compare the models with observations, they are useful to a better understanding of the physical mechanism responsible for the observed dust polarization power spectra in the \planck\ data. 

Our results support a scenario in which a large-scale feature of the magnetic field may produce the observed $T$-$B$ and $T$-$E$ correlations at large angular scales, which would otherwise be unexplained, although we cannot definitely conclude that the Galactic magnetic field structure within a few hundred pc from the Sun has a true helical component. We suggest that such large-scale structures of the magnetic field may be due to the characteristics of the local environment around the Sun, such as the expansion of the {\it local bubble} and its recently-modeled impact on the Galactic magnetic field \citep{Alves2018}. Future observational investigations of helical magnetic fields in the solar neighborhood may rely on the joint analysis of complementary magnetic-field tracers in the ISM, such as dust polarization and rotation measures (RM) with ancillary and novel large-scale data coming online \citep{Taylor2009,Shimwell2018,Tassis2018}. Despite the many caveats (i.e., the different gas phases probed by the two observational techniques) such analysis would allow one to attempt a 3D characterization of the Galactic magnetic-field structure combining the plane-of-the-sky magnetic field from dust polarization with the line-of-sight magnetic field from RM.

Helical turbulence in the ISM would possibly affect our latter results that mainly focus on the large-scale structure of the Galactic magnetic field around the Sun. \citet{Brandenburg2018} explore the link between fully helical MHD turbulence and the polarization power spectra, but it lacks the interplay with a mean field on large scales. 

In order to have a complete picture, it is necessary to examine the effect of turbulence on the helical Galactic magnetic field and to quantitatively compare the relative importance of turbulent and mean components in influencing the polarization power spectra and the cross-correlations. This can be done via self-consistent numerical MHD simulations,
but this is beyond the scope of the present work and constitutes a possible follow-up project.  

Finally, we also point out that, in spite of its astrophysical explanation, the helical component of the magnetic field that we introduced may be an interesting input for creating template model maps of dust polarization in the context of CMB foreground analyses. Because of the simple implementation of our model, the helical component of the field may be tuned to reach the level of $T$-$E$ and $T$-$B$ correlations in the data and simply added to present dust foreground models, similar to what was proposed by \citet{Vansyngel2017}.      

\section{Summary}\label{sec:conc}
We have presented toy models of helical magnetic fields in the ISM to gain insight into recent \planck\ observational results about Galactic dust polarization power spectra, or the positive $T$-$E$ and $T$-$B$ correlations, with particular focus on the latter. Since dust polarization depends on the morphology of the Galactic magnetic field, the characteristic property of $B$ modes of changing sign under parity transformations pushed us to explore the link of the $T$-$B$ correlation with magnetic helicity, which is as well an odd quantity under parity transformations.

The main results of our analysis are the following:
\begin{itemize}
\item the sign of magnetic helicity does not affect $E$ and $B$ mode power spectra, for isotropic magnetic-field configurations;
\item helical magnetic fields around interstellar filaments are not able to reproduce the $T$-$B$ correlation observed in the \planck\ data;
\item weak helical left-handed magnetic fields in the solar neighborhood (within a few hundred pc from us) can explain qualitatively the observed positive $T$-$E$ and $T$-$B$ correlations found at large angular scale in the \planck\ analyses.  
\end{itemize}
Our work represents a paradigm change in interpreting the dust polarization power spectra given so far. 
We propose a scenario in which the observed $T$-$E$ and $T$-$B$ correlations at large angular scales (for multipoles $l < 50$) are consequence of the large-scale structure of the interstellar magnetic field in the solar neighborhood, instead of small-scale MHD turbulent processes in the ISM.
Our results open interesting lines of research, such as the use of the $E$-$B$ polarization modes decomposition to characterize magnetic helicity in astrophysical environments; the study of the large-scale dynamics in the Milky Way around the Sun; the modeling of template maps of dust polarization for CMB foreground analyses.
\begin{acknowledgements}
First and foremost we would like to thank the anonymous referee for improving the clarity of the paper. We are grateful to Francois Boulanger for the encouraging conversations and his useful comments on the early versions of this work. We also thank Nicolas Ponthieu for teaching us how to play \verb|POKER|. 
Andrea Bracco, Simon Candelaresi, and Fabio Del Sordo acknowledge NORDITA for hospitality during the scientific program "{\it Phase transition in Astrophysics}", when the idea of this work was first conceived. Fabio Del Sordo acknowledges Artemis Del Sordo for useful discussions, and the Astrophysics group at University of Crete for warm hospitality. The work of Axel Brandenburg was supported by the National Science Foundation through the Astrophysics and Astronomy Grant Program (grant 1615100) and the University of Colorado through
the George Ellery Hale visiting faculty appointment.
Simon Candelaresi acknowledges support from the UK's
STFC (grant number ST/K000993)
\end{acknowledgements}

\bibliographystyle{aa}
\bibliography{aanda}

\begin{thebibliography}{57}
\expandafter\ifx\csname natexlab\endcsname\relax\def\natexlab#1{#1}\fi

\bibitem[{{Alves} {et~al.}(2018){Alves}, {Boulanger}, {Ferri{\`e}re}, \&
  {Montier}}]{Alves2018}
{Alves}, M.~I.~R., {Boulanger}, F., {Ferri{\`e}re}, K., \& {Montier}, L. 2018,
  \aap, 611, L5

\bibitem[{{Armstrong} {et~al.}(1995){Armstrong}, {Rickett}, \&
  {Spangler}}]{Armstrong1995}
{Armstrong}, J.~W., {Rickett}, B.~J., \& {Spangler}, S.~R. 1995, \apj, 443, 209

\bibitem[{{Bally} {et~al.}(1987){Bally}, {Langer}, {Stark}, \&
  {Wilson}}]{Bally1987}
{Bally}, J., {Langer}, W.~D., {Stark}, A.~A., \& {Wilson}, R.~W. 1987, \apjl,
  312, L45

\bibitem[{{BICEP2/Keck Collaboration} \& {Planck
  Collaboration}(2015)}]{BicepPlanck}
{BICEP2/Keck Collaboration} \& {Planck Collaboration}. 2015, Phys.\ Rev.\
  Lett., 114, 101301

\bibitem[{{Brandenburg} {et~al.}(2018){Brandenburg}, {Bracco}, {Kahniashvili},
  {Mandal}, {Roper Pol}, {Petrie}, \& {Singh}}]{Brandenburg2018}
{Brandenburg}, A., {Bracco}, A., {Kahniashvili}, T., {et~al.} 2018, ArXiv
  e-prints [\eprint[arXiv]{1807.11457}]

\bibitem[{{Brandenburg} \& {Lazarian}(2013)}]{Brandenburg2013}
{Brandenburg}, A. \& {Lazarian}, A. 2013, \ssr, 178, 163

\bibitem[{{Brandenburg} \& {Subramanian}(2005)}]{BS05c}
{Brandenburg}, A. \& {Subramanian}, K. 2005, Astronomische Nachrichten, 326,
  400

\bibitem[{{Caldwell} {et~al.}(2017){Caldwell}, {Hirata}, \&
  {Kamionkowski}}]{Caldwell2017}
{Caldwell}, R.~R., {Hirata}, C., \& {Kamionkowski}, M. 2017, Astrophys.\ J.,
  839, 91

\bibitem[{{Chandrasekhar} \& {Fermi}(1953)}]{Chandra1953}
{Chandrasekhar}, S. \& {Fermi}, E. 1953, Astrophys.\ J., 118, 113

\bibitem[{{Chepurnov} \& {Lazarian}(2010)}]{Chepurnov2010}
{Chepurnov}, A. \& {Lazarian}, A. 2010, \apj, 710, 853

\bibitem[{{Clark} {et~al.}(2015){Clark}, {Hill}, {Peek}, {Putman}, \&
  {Babler}}]{Clark2015}
{Clark}, S.~E., {Hill}, J.~C., {Peek}, J.~E.~G., {Putman}, M.~E., \& {Babler},
  B.~L. 2015, Phys.\ Rev.\ Lett., 115, 241302

\bibitem[{{Davis} \& {Greenstein}(1951)}]{Davis1951}
{Davis}, Jr., L. \& {Greenstein}, J.~L. 1951, Astrophys.\ J., 114, 206

\bibitem[{{Fiege} \& {Pudritz}(2000)}]{Fiege00}
{Fiege}, J.~D. \& {Pudritz}, R.~E. 2000, Astrophys.\ J., 544, 830

\bibitem[{{Fujimoto} \& {Miyamoto}(1969)}]{Fujimoto1969}
{Fujimoto}, M. \& {Miyamoto}, M. 1969, \pasj, 21, 194

\bibitem[{{Galloway} \& {Frisch}(1986)}]{Galloway1986}
{Galloway}, D. \& {Frisch}, U. 1986, Geophysical and Astrophysical Fluid
  Dynamics, 36, 53

\bibitem[{{Grain} {et~al.}(2012){Grain}, {Tristram}, \& {Stompor}}]{Grain2012}
{Grain}, J., {Tristram}, M., \& {Stompor}, R. 2012, \prd, 86, 076005

\bibitem[{{Hall}(1949)}]{Hall1949}
{Hall}, J.~S. 1949, Sci.\ J., 109, 166

\bibitem[{{Heiles}(1996)}]{Heiles1996a}
{Heiles}, C. 1996, in Astronomical Society of the Pacific Conference Series,
  Vol.~97, Polarimetry of the Interstellar Medium, ed. W.~G. {Roberge} \&
  D.~C.~B. {Whittet}, 457

\bibitem[{{HI4PI Collaboration} {et~al.}(2016){HI4PI Collaboration}, {Ben
  Bekhti}, {Fl{\"o}er}, {Keller}, {Kerp}, {Lenz}, {Winkel}, {Bailin},
  {Calabretta}, {Dedes}, {Ford}, {Gibson}, {Haud}, {Janowiecki}, {Kalberla},
  {Lockman}, {McClure-Griffiths}, {Murphy}, {Nakanishi}, {Pisano}, \&
  {Staveley-Smith}}]{HI4PI2016}
{HI4PI Collaboration}, {Ben Bekhti}, N., {Fl{\"o}er}, L., {et~al.} 2016, \aap,
  594, A116

\bibitem[{{Hiltner}(1949)}]{Hiltner1949}
{Hiltner}, W.~A. 1949, Sci.\ J., 109, 165

\bibitem[{{Hoang} {et~al.}(2018){Hoang}, {Cho}, \&
  {Lazarian}}]{HoangLazarian2018}
{Hoang}, T., {Cho}, J., \& {Lazarian}, A. 2018, Astrophys.\ J., 852, 129

\bibitem[{{Hoang} \& {Lazarian}(2016)}]{HoangLazarian2016}
{Hoang}, T. \& {Lazarian}, A. 2016, Astrophys.\ J., 831, 159

\bibitem[{{Hu} \& {White}(1997)}]{Hu1997}
{Hu}, W. \& {White}, M. 1997, Astrophys.\ J., 486, L1

\bibitem[{{Kahniashvili} {et~al.}(2014){Kahniashvili}, {Maravin},
  {Lavrelashvili}, \& {Kosowsky}}]{Kahniashvili2014}
{Kahniashvili}, T., {Maravin}, Y., {Lavrelashvili}, G., \& {Kosowsky}, A. 2014,
  \prd, 90, 083004

\bibitem[{{Kamionkowski} {et~al.}(1997){Kamionkowski}, {Kosowsky}, \&
  {Stebbins}}]{Kamionkowski1997}
{Kamionkowski}, M., {Kosowsky}, A., \& {Stebbins}, A. 1997, Phys.\ Rev.\ Lett.,
  78, 2058

\bibitem[{{Kandel} {et~al.}(2017){Kandel}, {Lazarian}, \&
  {Pogosyan}}]{Kandel2017}
{Kandel}, D., {Lazarian}, A., \& {Pogosyan}, D. 2017, \mnras, 472, L10

\bibitem[{{Kandel} {et~al.}(2018){Kandel}, {Lazarian}, \&
  {Pogosyan}}]{Kandel2018}
{Kandel}, D., {Lazarian}, A., \& {Pogosyan}, D. 2018, \mnras, 478, 530

\bibitem[{Kritsuk {et~al.}(2018)Kritsuk, Flauger, \& Ustyugov}]{Kritsuk2017}
Kritsuk, A.~G., Flauger, R., \& Ustyugov, S.~D. 2018, Phys.\ Rev.\ Lett., 121,
  021104

\bibitem[{{Lazarian} \& {Hoang}(2007)}]{Lazarian2007}
{Lazarian}, A. \& {Hoang}, T. 2007, \mnras, 378, 910

\bibitem[{{Lazarian} \& {Yuen}(2018)}]{Lazarian2018}
{Lazarian}, A. \& {Yuen}, K.~H. 2018, \apj, 853, 96

\bibitem[{{Mathewson}(1968)}]{Mathewson68}
{Mathewson}, D.~S. 1968, \apjl, 153, L47

\bibitem[{{Matthews} \& {Wilson}(2002)}]{Matthews2002}
{Matthews}, B.~C. \& {Wilson}, C.~D. 2002, \apj, 571, 356

\bibitem[{{Planck Collaboration Int. XIX}(2015)}]{PIPXIX2015}
{Planck Collaboration Int. XIX}. 2015, \aap, 576, A104

\bibitem[{{Planck Collaboration Int. XLIV}(2016)}]{PIPXLIV2016}
{Planck Collaboration Int. XLIV}. 2016, \aap, 596, A105

\bibitem[{{Planck Collaboration Int. XX}(2015)}]{PIPXX2015}
{Planck Collaboration Int. XX}. 2015, \aap, 576, A105

\bibitem[{{Planck Collaboration Int. XXX}(2016)}]{planck2014-XXX}
{Planck Collaboration Int. XXX}. 2016, \aap, 586, A133

\bibitem[{{Planck Collaboration Int. XXXII}(2016)}]{PIPXXXII}
{Planck Collaboration Int. XXXII}. 2016, \aap, 586, A135

\bibitem[{{Planck Collaboration Int. XXXVIII}(2016)}]{planck2015-XXXVIII}
{Planck Collaboration Int. XXXVIII}. 2016, \aap, 586, A141

\bibitem[{{Planck Collaboration results. I.}(2016)}]{PlanckI2016}
{Planck Collaboration results. I.} 2016, \aap, 594, A1

\bibitem[{{Planck Collaboration results. XI.}(2018)}]{PIPXI2018}
{Planck Collaboration results. XI.} 2018, ArXiv e-prints
  [\eprint[arXiv]{1801.04945}]

\bibitem[{{Poidevin} {et~al.}(2011){Poidevin}, {Bastien}, \&
  {Jones}}]{Poidevin2011}
{Poidevin}, F., {Bastien}, P., \& {Jones}, T.~J. 2011, \apj, 741, 112

\bibitem[{{Ponthieu} {et~al.}(2011){Ponthieu}, {Grain}, \&
  {Lagache}}]{Ponthieu2011}
{Ponthieu}, N., {Grain}, J., \& {Lagache}, G. 2011, \aap, 535, A90

\bibitem[{{Robitaille} {et~al.}(2017){Robitaille}, {Scaife}, {Carretti},
  {Gaensler}, {McEwen}, {Leistedt}, {Haverkorn}, {Bernardi}, {Kesteven},
  {Poppi}, \& {Staveley-Smith}}]{Robitaille2017}
{Robitaille}, J.-F., {Scaife}, A.~M.~M., {Carretti}, E., {et~al.} 2017, \mnras,
  468, 2957

\bibitem[{{Seljak} \& {Zaldarriaga}(1997)}]{Seljak1997}
{Seljak}, U. \& {Zaldarriaga}, M. 1997, Phys.\ Rev.\ Lett., 78, 2054

\bibitem[{{Shimwell} {et~al.}(2018){Shimwell}, {Tasse}, {Hardcastle}, {Mechev},
  {Williams}, {Best}, {R{\"o}ttgering}, {Callingham}, {Dijkema}, {de Gasperin},
  {Hoang}, {Hugo}, {Mirmont}, {Oonk}, {Prandoni}, {Rafferty}, {Sabater},
  {Smirnov}, {van Weeren}, {White}, {Atemkeng}, {Bester}, {Bonnassieux},
  {Br{\"u}ggen}, {Brunetti}, {Chy{\.z}y}, {Cochrane}, {Conway}, {Croston},
  {Danezi}, {Duncan}, {Haverkorn}, {Heald}, {Iacobelli}, {Intema}, {Jackson},
  {Jamrozy}, {Jarvis}, {Lakhoo}, {Mevius}, {Miley}, {Morabito}, {Morganti},
  {Nisbet}, {Orr{\'u}}, {Perkins}, {Pizzo}, {Schrijvers}, {Smith}, {Vermeulen},
  {Wise}, {Alegre}, {Bacon}, {van Bemmel}, {Beswick}, {Bonafede}, {Botteon},
  {Bourke}, {Brienza}, {Calistro Rivera}, {Cassano}, {Clarke}, {Conselice},
  {Dettmar}, {Drabent}, {Dumba}, {Emig}, {En{\ss}lin}, {Ferrari}, {Garrett},
  {G{\'e}nova-Santos}, {Goyal}, {G{\"u}rkan}, {Hale}, {Harwood}, {Heesen},
  {Hoeft}, {Horellou}, {Jackson}, {Kokotanekov}, {Kondapally},
  {Kunert-Bajraszewska}, {Mahatma}, {Mahony}, {Mandal}, {McKean}, {Merloni},
  {Mingo}, {Miskolczi}, {Mooney}, {Nikiel-Wroczy{\'n}ski}, {O'Sullivan},
  {Quinn}, {Reich}, {Roskowi{\'n}ski}, {Rowlinson}, {Savini}, {Saxena},
  {Schwarz}, {Shulevski}, {Sridhar}, {Stacey}, {Urquhart}, {van der Wiel},
  {Varenius}, {Webster}, \& {Wilber}}]{Shimwell2018}
{Shimwell}, T.~W., {Tasse}, C., {Hardcastle}, M.~J., {et~al.} 2018, ArXiv
  e-prints [\eprint[arXiv]{1811.07926}]

\bibitem[{{Shukurov} {et~al.}(2006){Shukurov}, {Sokoloff}, {Subramanian}, \&
  {Brandenburg}}]{SSSB06}
{Shukurov}, A., {Sokoloff}, D., {Subramanian}, K., \& {Brandenburg}, A. 2006,
  \aap, 448, L33

\bibitem[{{Sur} {et~al.}(2007){Sur}, {Shukurov}, \& {Subramanian}}]{Sur2007}
{Sur}, S., {Shukurov}, A., \& {Subramanian}, K. 2007, \mnras, 377, 874

\bibitem[{{Tahani} {et~al.}(2018){Tahani}, {Plume}, {Brown}, \&
  {Kainulainen}}]{Tahani2018}
{Tahani}, M., {Plume}, R., {Brown}, J.~C., \& {Kainulainen}, J. 2018, \aap,
  614, A100

\bibitem[{{Tashiro} {et~al.}(2014){Tashiro}, {Chen}, {Ferrer}, \&
  {Vachaspati}}]{Tashiro2014}
{Tashiro}, H., {Chen}, W., {Ferrer}, F., \& {Vachaspati}, T. 2014, \mnras, 445,
  L41

\bibitem[{{Tassis} \& {Pavlidou}(2015)}]{TassisPavlidou2015}
{Tassis}, K. \& {Pavlidou}, V. 2015, \mnras, 451, L90

\bibitem[{{Tassis} {et~al.}(2018){Tassis}, {Ramaprakash}, {Readhead}, {Potter},
  {Wehus}, {Panopoulou}, {Blinov}, {Eriksen}, {Hensley}, {Karakci},
  {Kypriotakis}, {Maharana}, {Ntormousi}, {Pavlidou}, {Pearson}, \&
  {Skalidis}}]{Tassis2018}
{Tassis}, K., {Ramaprakash}, A.~N., {Readhead}, A.~C.~S., {et~al.} 2018, ArXiv
  e-prints [\eprint[arXiv]{1810.05652}]

\bibitem[{{Taylor} {et~al.}(2009){Taylor}, {Stil}, \& {Sunstrum}}]{Taylor2009}
{Taylor}, A.~R., {Stil}, J.~M., \& {Sunstrum}, C. 2009, \apj, 702, 1230

\bibitem[{{Toci} \& {Galli}(2015)}]{Toci2015}
{Toci}, C. \& {Galli}, D. 2015, \mnras, 446, 2118

\bibitem[{{Vansyngel} {et~al.}(2017){Vansyngel}, {Boulanger}, {Ghosh},
  {Wandelt}, {Aumont}, {Bracco}, {Levrier}, {Martin}, \&
  {Montier}}]{Vansyngel2017}
{Vansyngel}, F., {Boulanger}, F., {Ghosh}, T., {et~al.} 2017, \aap, 603, A62

\bibitem[{{Vishniac} \& {Cho}(2001)}]{Vishniac2001}
{Vishniac}, E.~T. \& {Cho}, J. 2001, \apj, 550, 752

\bibitem[{{Zaldarriaga}(2001)}]{Zaldarriaga2001}
{Zaldarriaga}, M. 2001, \prd, 64, 103001

\bibitem[{{Zaldarriaga} \& {Seljak}(1997)}]{Zaldarriaga1997}
{Zaldarriaga}, M. \& {Seljak}, U. 1997, \prd, 55, 1830

\end{thebibliography}
\appendix
\section{E and B modes power spectra reconstruction in a flat sky}\label{app:eb}

In this Appendix we detail how we compute the $E$ and $B$ modes in the small-scale limit on a flat sky. This step is key since our methodology mostly produces 2D synthetic maps of the Stokes parameters. We refer to the discussion in Sect.~V of \citet{Zaldarriaga1997} and we report in the following the two main equations we use to perform the $E$-$B$ decomposition in the Fourier space,
\begin{align}\label{eq:eb2D}
\tilde{E}_{k} &= \tilde{Q}_{k}\, \cos{2\phi_{k}} + \tilde{U}_{k}\, \sin{2\phi_{k}}\\
\tilde{B}_{k} &= -\tilde{Q}_{k}\, \sin{2\phi_{k}} + \tilde{U}_{k}\, \cos{2\phi_{k}},\nonumber
\end{align}
where the "$\sim$" sign indicates the Fourier transforms of the corresponding quantities and $\phi_k$ is the orientation of the wave-vector {$\bm{k}$} in the Fourier space. We use the above equations only in the case of periodic boundary conditions in the boxes (see discussion in \citet{Ponthieu2011} for non-periodic boundary conditions).
\begin{figure}[htbp] 
   \centering
   \includegraphics[width=0.5\textwidth]{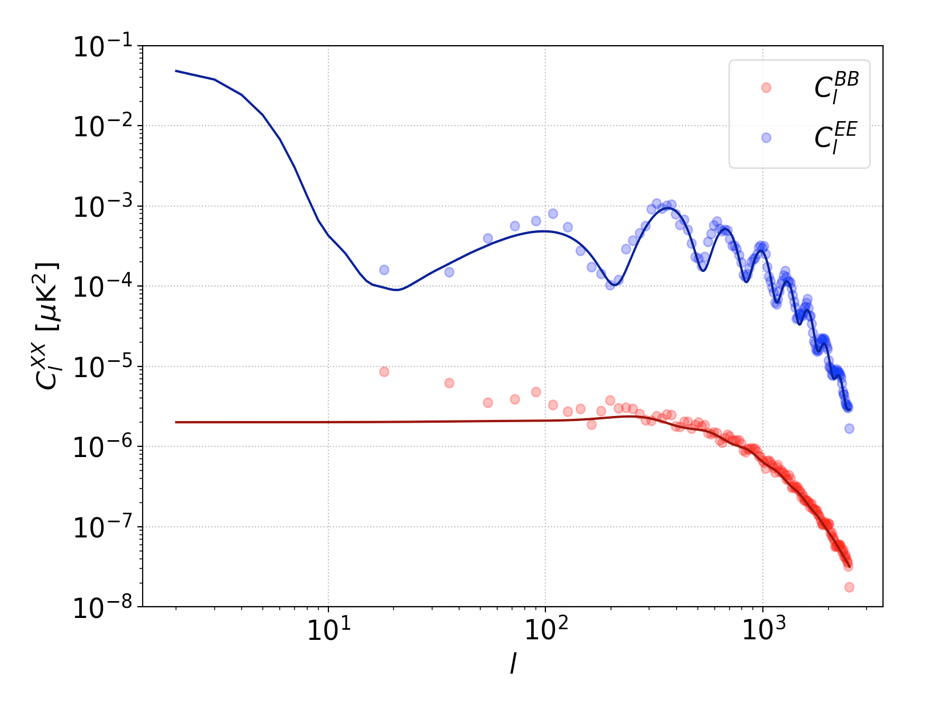}
   \caption{Validation of the $E$-$B$ modes decomposition. Solid lines are input CMB power spectra, while colored dots are the output ones obtained with our algorithm.}
   \label{fig:cmb1}
\end{figure} \\
In order to validate our algorithms we show an example based on the best-fit $\Lambda$CDM CMB $E$ and $B$ power spectra from the \planck\ PR3 baseline (\verb|COM_PowerSpect_CMB-base-plikHM-TTTEEE-lowl-lowE-lensing-minimum-theory_R3.01.txt| in \url{http://pla.esac.esa.int/pla/#cosmology}). In Fig.~\ref{fig:cmb1} the blue and red solid lines correspond to the $E$ and $B$ input power spectra, which can be downloaded from the aforementioned link, while the colored dots represent the output-reconstructed spectra using our methodology. In practice given the input $E$ and $B$ mode power spectra, we produce a random realization of $Q$ and $U$ maps of the CMB (see Fig.~\ref{fig:cmb2}) with the \verb|IDL| routine \verb|maps_iqu2cls.pro| of the \verb|POKER| package \citep{Ponthieu2011} and we use Eq.~(\ref{eq:eb2D}) to extract the output data points shown in Fig.~\ref{fig:cmb1}. We are able to retrieve the input power spectra with high accuracy. Notice that: i) we convert $k$ to multipoles $l$; ii) the very large scales are the most affected by cosmic variance. 
\begin{figure}[htbp] 
   \centering
   \includegraphics[width=0.5\textwidth]{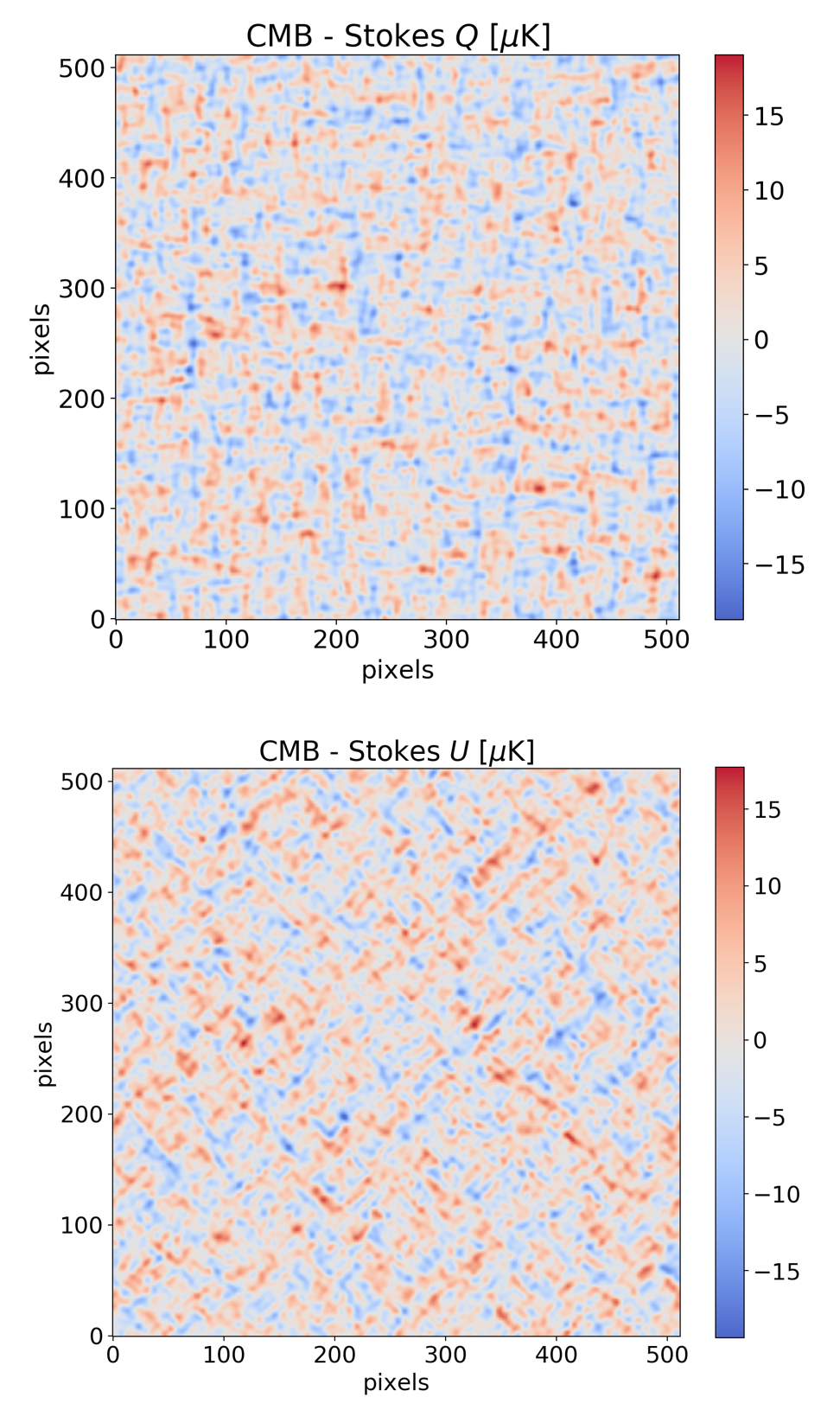}
   \caption{Random realization of the $Q$ and $U$ CMB Stokes maps from the power spectra shown with solid lines in Fig.~\ref{fig:cmb1}. The simulated field of view is $10^{\circ}\times 10^{\circ}$. These are the maps we use to validate our method.}
   \label{fig:cmb2}
\end{figure}
\end{document}